\documentstyle [aps,preprint]{revtex}
\tightenlines
\begin{document}
\def\fnote#1#2{
\begingroup\def\thefootnote{#1}\footnote{#2}\addtocounter{footnote}{-1}
\endgroup}
\def\dslash{\not{\hbox{\kern-2pt $\partial$}}}
\def\eslash{\not{\hbox{\kern-2pt $\epsilon$}}}
\def\Dslash{\not{\hbox{\kern-4pt $D$}}}
\def\Aslash{\not{\hbox{\kern-4pt $A$}}}
\def\Qslash{\not{\hbox{\kern-4pt $Q$}}}
\def\Wslash{\not{\hbox{\kern-4pt $W$}}}
\def\pslash{\not{\hbox{\kern-2.3pt $p$}}}
\def\kslash{\not{\hbox{\kern-2.3pt $k$}}}
\def\qslash{\not{\hbox{\kern-2.3pt $q$}}}
\def\np#1{{\sl Nucl.~Phys.~\bf B#1}}
\def\pl#1{{\sl Phys.~Lett.~\bf B#1}}
\def\pr#1{{\sl Phys.~Rev.~\bf D#1}}
\def\prl#1{{\sl Phys.~Rev.~Lett.~\bf #1}}
\def\cpc#1{{\sl Comp.~Phys.~Comm.~\bf #1}}
\def\cmp#1{{\sl Commun.~Math.~Phys.~\bf #1}}
\def\anp#1{{\sl Ann.~Phys.~(NY) \bf #1}}
\def\etal{{\em et al.}}
\def\half{{\textstyle{1\over2}}}
\def\be{\begin{equation}}
\def\ee{\end{equation}}
\def\ba{\begin{array}}
\def\ea{\end{array}}
\def\tr{{\rm tr}}
\def\Tr{{\rm Tr}}
\title{Scalar absorption by spinning D3-branes
\thanks{Research
supported by the DoE under grant DE--FG05--91ER40627.}}
\author{
George Siopsis
\fnote{\dagger}{\tt gsiopsis@utk.edu}}
\address{Department of Physics and Astronomy, \\
The University of Tennessee, Knoxville, TN 37996--1200.\\
}
\date{August 1998}
\preprint{UTHET--99--0801}
\maketitle
\begin{abstract}
We discuss absorption of scalars by a distribution of spinning D3-branes.
The D3-branes are multi-center solutions of supergravity theory.
We solve the wave equation in various cases of supergravity backgrounds
in which the equation becomes separable.
We show that the absorption coefficients exhibit a universal behavior as
functions of the angular momentum quantum number and the Hawking temperature.
This behavior is similar to the form of the gray-body factors one
encounters in scattering by black-holes.
Our discussion includes the problematic case of spherically symmetric distributions
of D-branes, where resonances arise. We obtain the same universal
form for the absorption coefficients,
if the region enclosed by the D-branes is excluded from physical considerations.
Non-extremal D-branes are also discussed.
The results are similar to the extremal cases, albeit at half the Hawking temperature.
We speculate that new degrees of freedom enter as one moves away from extremality.
\end{abstract}
\renewcommand\thepage{}\newpage\pagenumbering{arabic}

\section{Introduction}

There exists ample evidence
of an exact correspondence~\cite{bib1,bib2,bib3} between ${\cal N} =4$
four-dimensional supersymmetric $SU(N)$ Yang-Mills theory in the large-$N$ limit and string theory in
a supergravity background representing a collection of D3-branes whose near-horizon geometry is the product of
an Anti-deSitter space and a five-dimensional sphere
($AdS_5\times S^5$)~\cite{bib1,bib2,bib3}.
This correspondence enables us to calculate correlation functions as well
as various thermodynamic properties of super Yang-Mills theories in the
large-$N$ limit using results from supergravity~\cite{bib4,bib5,bib6,bib7}. Exact results are obtained
primarily due to the superconformal invariance of super Yang-Mills theory.

The metric for a stack of coincident D3-branes is
\be\label{deka}
ds^2 = {1\over\sqrt H} \; \left( -dt^2+dx_1^2+dx_2^2+dx_3^2 \right)
+ \sqrt H\; \left( dy_1^2+\dots + dy_6^2  \right)
\ee
where
\be\label{endeka}
H = 1 + {R^4\over |{\bf y}|^4} \;,\quad\quad |{\bf y}|^2 = dy_1^2+\dots + dy_6^2
\ee
and the dilaton field is constant. Near the horizon ($|{\bf y}|\to 0$), one may drop
the constant term in the harmonic function $H$ and the resulting metric
describes the space $AdS_5\times S^5$. The $AdS_5$ throat size, $R$, is also the
curvature of $S^5$. We obtain exact superconformal
invariance on the boundary of the AdS space.

This is a special solution of the supergravity field equations. It has zero
temperature and is maximally supersymmetric. If our goal is to understand
quantum gravity, we need to study a larger set of backgrounds that possess
less symmetry and contain the special case~(\ref{deka}) with $H$ given by~(\ref{endeka}) as a limit point.
Such studies have already appeared in the literature, starting with linearized
perturbations around the special solution~\cite{bib7ab}
and including solutions of the full non-linear
field equations~\cite{bib7a,bib7aa}. Of particular importance are solutions that represent
a collection of branes (multi-center). In this case, the harmonic function
becomes
\be\label{dodeka}
H = 1 + R^4\; \int d^6 y' {\sigma ({\bf y'})\over |{\bf y} - {\bf y'}|^4}
\;, \quad\quad \int d^6 y' \sigma ({\bf y'}) = 1
\ee
where $\sigma ({\bf y})$ describes the distribution of branes, which can be
discreet or continuous. These D-branes correspond to a broken phase of the
super Yang-Mills theory where certain operators develop non-vanishing
vacuum expectation values. This is the Coulomb branch of the gauge theory,
because the remaining massless bosons mediate long-range Coulomb
interactions. Superconformal symmetry is broken. The space of these solutions
is the moduli space of the Yang-Mills theory. In this language, the special
solution~(\ref{deka}) with $H$ given by~(\ref{endeka}) corresponds to the origin of moduli space.

From the supergravity point of view, even Eq.~(\ref{deka}) with $H$ given by~(\ref{dodeka}) is but a special class of solutions (extreme solutions). A larger class
of non-extreme solutions has been found~\cite{bib7aa,bib7b,bib7c}. It might be of interest to study
this larger class of solutions, even though one expects that they correspond
to unstable states in the Hilbert space of the gauge theory. They are all
finite temperature configurations and they might shed some light on the
thermodynamic properties of gauge theories, such as phase transitions~\cite{bib8}.

A useful tool in these investigations is the study of interactions of branes
with external probes~\cite{bib8a,bib8b,bib9,bib9a}. In particular, the absorption cross-section for a scalar
in an AdS background has been shown to agree with the one obtained from
superconformal field theory. This agreement has been shown to be exact in the
low energy (for the scalar) limit and for all partial waves of the scalar
field~\cite{bib8a}. Extensions to higher energies have also been considered~\cite{bib9}. In the more
general case of non-coincident D-branes, such calculations are considerably
more involved, because the wave equation becomes non-separable. Certain
distributions of D-branes have been discussed for the scattering of $s$-waves.

It should also be mentioned that the case of a spherically symmetric distribution
of D-branes has presented a puzzle~\cite{bib10}. In this case, the incident scalar field
exhibits resonant behavior at certain values of its energy. These special
frequencies extend all the way to infinity and are multiples of $\ell/R^2$,
where $\ell$ is the size of the distribution of the D-branes. Thus, they
all go to zero as we approach the AdS limit ($\ell\to 0$), but it is not
clear how that limit is to be described. These resonances arise if one
allows for reflection of the incident wave off of the D-branes without
accounting for absorption.

Here, we extend the study of absorption of scalars by a distribution of
D3-branes to include a larger set of supergravity backgrounds than previously
considered and arbitrary
partial waves. We solve the
wave equation in the respective backgrounds for various D-brane
distributions in the extremal limit. We also extend the analysis to
the case of a non-extremal D3-brane. In general, the waves become singular
at the positions of the D-branes.
We find that the absorption coefficients exhibit a universal
behavior similar to the form of the gray-body factors in the case of
black-hole scattering~\cite{bib9b}.

Our calculations include the troublesome spherically
symmetric D3-brane distribution where resonances arise for an infinite number
of frequencies of the incident wave~\cite{bib10}.
It is easy to understand the origin of these resonances. If the D-branes cover a closed surface in the
transverse space (with coordinates $y_1,\dots,y_6$), then the possibility
arises of multiple reflections of the incident wave off of the D-branes.
We demonstrate this for a spherical shell of D-branes as well as a long needle
(cylindrical symmetry). The waves are no longer singular
on the D-brane surfaces. Then it is natural to ask what type
of discontinuity the D-branes should impose on the incident wavefunction.
This question will probably be settled by a conformal field theoretical
calculation. Here we show that if reflection is forbidden, the absorption
coefficients exhibit the same behavior as in the ``healthier" cases, where no resonances
arise ({\em i.e.,}~when the D-branes are distributed over an open surface in the
transverse space). There is no reflection if the region enclosed by the
D-brane distribution is excluded from physical considerations. Thus, it appears
that the Schwarzschild-like coordinates one works with in supergravity are more
physically relevant than the $y_i$ ($i=1,\dots,6$) coordinates of the transverse
space in Eq.~(\ref{deka}).

Our discussion is organized as follows.
In Section~\ref{sec2}, we introduce the metric for a
collection of spinning D3-branes in the extremal limit and solve the
wave equation in the respective backgrounds.
In Section~\ref{sec3}, we extend the analysis to
the case of a non-extremal D3-brane. The results are similar to the
extremal cases, albeit at {\em half} the Hawking temperature.
Our conclusions are summarized in Section~\ref{sec4}.

\section{Extremal D3-branes}\label{sec2}

In this Section, we solve the wave equation for a scalar field in a supergravity
background representing a distribution of D3-branes in the extremal limit.
For completeness, we start with the metric for non-extremal spinning D3-branes and
then take the extremal limit. In this limit, the branes are no longer spinning,
but they settle to a state which is distinct from the AdS limit. These new vacua
are stable due to the existence of a chemical potential. They are distinguished
by quantum numbers that correspond to the angular momentum in the non-extremal
regime~\cite{bib7aa}.

The metric for a general distribution of spinning D3-branes in ten dimensions
is~\cite{bib7aa,bib7b,bib7c}\fnote{\clubsuit}{We have fixed various typographical errors in
refs.~\cite{bib7aa,bib7b,bib7c}}
$$ds^2 = {1\over\sqrt H} \left( - (1- fr_0^4/r^4) dt^2 + dx_1^2+dx_2^2+dx_3^2 \right)
+ \sqrt H f^{-1} {dr^2\over \lambda - r_0^4/r^4}$$
$$+ \sqrt H r^2 \left( \zeta d\theta^2 + \zeta' \cos^2\theta d\psi^2 -
{\ell_2^2-\ell_3^2\over 2r^2} \sin (2\theta) \sin (2\psi) d\theta d\psi\right)$$
$$- f {2r_0^4\cosh\gamma\over r^4} \; \sqrt H\; (\ell_1\sin^2\theta d\phi_1 +
\ell_2\cos^2\theta \sin^2\psi d\phi_2 + \ell_3\cos^2\theta \cos^2\psi d\phi_3)\; dt$$
$$+ f {r_0^4\over r^4} \sqrt H\; (\ell_1\sin^2\theta d\phi_1 +
\ell_2\cos^2\theta \sin^2\psi d\phi_2 + \ell_3\cos^2\theta \cos^2\psi d\phi_3)^2$$
\be\label{miden}
+ \sqrt H r^2 \left[ \left(1+{\ell_1^2\over r^2} \right) \sin^2\theta d\phi_1^2
+\left(1+{\ell_2^2\over r^2} \right) \cos^2\theta \sin^2\psi d\phi_2^2
+\left(1+{\ell_3^2\over r^2} \right)\cos^2\theta \cos^2\psi d\phi_3^2 \right]
\ee
where
\be
H = 1+f {r_0^4\sinh^2\gamma\over r^4} \quad\quad f^{-1} = \lambda
\left( {\sin^2\theta\over 1+{\ell_1^2\over r^2}} + {\cos^2\theta \sin^2\psi
\over 1+{\ell_2^2\over r^2}} + {\cos^2\theta \cos^2\psi \over 1+{\ell_3^2\over r^2}}
\right)
\ee
\be
\lambda = \left(1+{\ell_1^2\over r^2} \right) \left(1+{\ell_2^2\over r^2} \right)
\left(1+{\ell_3^2\over r^2} \right)
\ee
\be\label{dekatria}
\zeta = 1+ {\ell_1^2\cos^2\theta + \ell_2^2\sin^2\theta\sin^2\psi
+\ell_3^2 \sin^2\theta\cos^2\psi \over r^2} \quad\quad
\zeta' = 1+ {\ell_2^2\cos^2\psi
+\ell_3^2 \sin^2\psi \over r^2}
\ee
and the charge of the branes is
\be
R^4 = {1\over 2}\; r_0^4 \sinh (2\gamma)
\ee
The horizon is the root of $\lambda - r_0^4/r^4 = 0$. The parameters $\ell_i$
($i=1,2,3$) are the angular momentum quantum numbers representing rotation around
axes in three distinct planes, respectively, in the six-dimensional transverse space.

In the extremal limit, the horizon shrinks to zero ($r_0\to 0$) and also
$\gamma \to\infty$, so that the charge $R^4$ remains finite. The angular momenta
also vanish and we obtain a static configuration. These configurations are
still described by the three angular momentum quantum numbers. They are at
finite temperature.
The metric in the extreme limit becomes
$$ds^2 = {1\over\sqrt H} \left( - dt^2 + dx_1^2+dx_2^2+dx_3^2 \right)
+ \sqrt H f^{-1} {dr^2\over \lambda}$$
$$+ \sqrt H r^2 \left( \zeta d\theta^2 + \zeta' \cos^2\theta d\psi^2 -
{\ell_2^2-\ell_3^2\over 2r^2} \sin (2\theta) \sin (2\psi) d\theta d\psi\right)$$
\be
+ \sqrt H r^2 \left[ \left(1+{\ell_1^2\over r^2} \right) \sin^2\theta d\phi_1^2
+\left(1+{\ell_2^2\over r^2} \right) \cos^2\theta \sin^2\psi d\phi_2^2
+\left(1+{\ell_3^2\over r^2} \right)\cos^2\theta \cos^2\psi d\phi_3^2 \right]
\ee
where
\be
H = 1+f {R^4\over r^4}
\ee
and the other functions, $f, \lambda, \zeta, \zeta'$ are still given by Eq.~(\ref{dekatria}). It can be shown that this metric is equivalent to the
multi-center form~(\ref{deka}) with $H$ given by~(\ref{dodeka}) through the
following transformation~\cite{bib7aa,bib11}
\be\label{cootr}
\begin{array}{l}
y_1 = \sqrt{r^2 + \ell_1^2} \sin\theta \cos\phi_1\\
y_2 = \sqrt{r^2 + \ell_1^2} \sin\theta \sin\phi_1\\
y_3 = \sqrt{r^2 + \ell_2^2} \cos\theta \sin\psi \cos\phi_2\\
y_4 = \sqrt{r^2 + \ell_2^2} \cos\theta \sin\psi \sin\phi_2\\
y_5 = \sqrt{r^2 + \ell_3^2} \cos\theta \cos\psi \cos\phi_3\\
y_6 = \sqrt{r^2 + \ell_3^2} \cos\theta \cos\psi \sin\phi_3
\end{array}
\ee
The wave equation in a general background is complicated. We will therefore
restrict attention to the special case $\ell_2 = \ell_3$ (``cylindrical" symmetry). We do not expect our conclusions to change in the more general case,
although it would take considerably more effort to prove it.
The metric becomes
$$ds^2 = {1\over\sqrt H} \left( - dt^2 + dx_1^2+dx_2^2+dx_3^2 \right)
+ \sqrt H \left(1+{\ell_2^2\over r^2} \right)\zeta {dr^2\over \lambda}
$$
\be
+ \sqrt H r^2 \left[ \zeta d\theta^2 + \left(1+{\ell_1^2\over r^2} \right) \sin^2\theta d\phi_1^2
+\left(1+{\ell_2^2\over r^2} \right) \cos^2\theta (d\psi^2+\sin^2\psi d\phi_2^2
+ \cos^2\psi d\phi_3^2) \right]
\ee
where
\be\label{ena1}
H = 1+ {R^4\over \left(1+{\ell_2^2\over r^2} \right)\zeta r^4}
\quad\quad
\lambda = \left(1+{\ell_1^2\over r^2} \right)\left(1+{\ell_2^2\over r^2} \right)^2
\quad\quad
\zeta = 1+ {\ell_1^2\cos^2\theta + \ell_2^2\sin^2\theta \over r^2}
\ee
Notice that the various functions, $H, \lambda, \zeta$, comprising the metric tensor, are functions of $r, \theta$
only. The ten-dimensional wave equation for a scalar field,
\be
\partial_A \sqrt{-g} g^{AB} \partial_B \Phi = 0
\ee
becomes separable for fields that are independent of the angular variables $\psi$, $\phi_i$
($i=1,2,3$).
Indeed, for a field of momentum $k_\mu$ and mass $m^2 = k_\mu k^\mu$,
\be
\Phi (x^\mu\, ;\, r\,,\, \theta) = e^{ik\cdot x}\; \Psi (r\,,\, \theta)
\ee
after some algebra, we obtain
\be
{1\over r^3\left(1+{\ell_2^2\over r^2} \right)} \partial_r \left(\lambda r^5 \partial_r \Psi\right)
+ m^2 r^2\Psi + {m^2R^4\over r^2
\left(1+{\ell_2^2\over r^2} \right)} \Psi - (\hat L^2 -m^2\ell_1^2\cos^2\theta-m^2\ell_2^2\sin^2\theta) \Psi =0
\ee
We will solve this equation in the limit where the mass is small
compared with the AdS curvature, and the angular momenta are also small,
\be
m R \ll 1 \; ,\quad\quad \ell_i \lesssim mR^2 \quad (i=1,2)
\ee
In this limit, the terms proportional to the angular momentum components,
$\ell_i$ ($i=1,2$), can
be dropped. Indeed, their contribution is
$m^2\ell_i^2 \ll m^2R^2 \ll 1$.
Therefore, they are small compared to the angular momentum ($\hat L^2$ term)
contribution.
The wave equation becomes
\be
{1\over r^3\left(1+{\ell_2^2\over r^2} \right)} \partial_r \left(\lambda r^5 \partial_r \Psi\right)
+ m^2r^2\Psi + {m^2R^4\over r^2
\left(1+{\ell_2^2\over r^2} \right)} \Psi - \hat L^2 \Psi =0
\ee
The eigenvalues of $\hat L^2$ are $j(j+4)$. Therefore, the radial part
of the wave equation is
\be
{1\over r^3\left(1+{\ell_2^2\over r^2} \right)} \left(\lambda r^5 \Psi'\right)'
+m^2 r^2\Psi + {m^2R^4\over r^2
\left(1+{\ell_2^2\over r^2} \right)} \Psi -  j(j+4)\Psi =0
\ee
We will solve this equation in two regimes, $r \gg mR^2$ and $r\ll 1/m$,
and then match the respective expressions asymptotically.

For $r \gg mR^2$, we obtain
\be
{1\over r^3} \left( r^5 \Psi'\right)'
+ m^2r^2\Psi -  j(j+4)\Psi =0
\ee
whose solution is
\be
\Psi = {1\over r^2} \; J_{j+2} (m r)
\ee
where we dropped the solution which is not regular at small $r$. The normalization is arbitrary, since we only care about ratios of fluxes.
At small $r$, the solution
behaves as
\be\label{tria}
\Psi \sim {m^2\over 4(j+2)!} \; \left( {m r\over 2} \right)^j
\ee
In the regime of small $r$ ($m r \ll 1$), the wave equation becomes
\be\label{duo1}
{1\over r^3\left(1+{\ell_2^2\over r^2} \right)} \left(\lambda r^5 \Psi'\right)'
 + {m^2R^4\over r^2
\left(1+{\ell_2^2\over r^2} \right)} \Psi -  j(j+4)\Psi =0
\ee
To solve this equation, we distinguish between three cases, in which
one, two or three components of the angular momentum are non-vanishing, respectively.

\subsection{One-component angular momentum}

The simplest case is the one where $\ell_2 = \ell_3 = 0$.
The D-branes are a limiting configuration of non-extremal branes spinning
around an axis in the plane defined by the coordinates $y_1, y_2$ in the
transverse space ({\em cf.}~Eq.~(\ref{cootr})). They are uniformly distributed on
a disk of radius $\ell_1$ in this plane~\cite{bib7a,bib7aa}. To see this, note that
the harmonic function $H$ (Eq.~(\ref{ena1}) can be written as
\be
H = 1+ {R^4\over \zeta r^4} = 1+ {R^4\over (r^2+\ell_1^2\cos^2\theta)r^2}
\ee
The D-branes are in the region bounded by the $r=0$ surface, which is a disk of
radius $\ell_1$ in the plane $y_3=\dots = y_6=0$ (because of Eq.~(\ref{cootr})).
Define
\be
y_{||}^2 = y_1^2+y_2^2 = (r^2+\ell_1^2) \sin^2\theta \;, \quad\quad
y_\perp^2 = y_3^2+\dots + y_6^2 = r^2\cos^2\theta
\ee
in terms of
which $H$ becomes
\be
H \approx 1+ {R^4\over \ell_1^2 y_\perp^2}
\ee
as $r\to 0$. The density of D-branes is therefore independent of $y_{||}$
and $\sigma = {1\over \pi\ell_1^2}$, {\em i.e.,} the D-branes are uniformly distributed on a disk of radius $\ell_1$ in the $y_{||}=0$ plane.

The wave equation for small $r$ (Eq.~(\ref{duo1})) is
\be
{1\over r^3} \left(\lambda r^5 \Psi'\right)'
 + {R^4m^2\over r^2} \Psi -  j(j+4)\Psi =0
\ee
where $\lambda = 1+{\ell_1^2\over r^2}$.
To solve this equation, change variables to $u = 1/\lambda$. Then
\be
(1-u) u^2 {d^2\Psi\over du^2} + u(2-u) {d\Psi\over du} + {m^2R^4\over 4\ell_1^2} \Psi - {j(j+4)u\over
4 (1-u)} \Psi =0
\ee
Next, we need to control the behavior at the singular points $u=0,1$.
As $u\to 0$, we obtain
\be\label{tria1}
u^2 {d^2\Psi\over du^2} + 2u {d\Psi\over du} +
{m^2R^4\over 4\ell_1^2} \Psi - {j(j+4)\over 4} u \Psi = 0
\ee
Assuming $\Psi \sim u^a$, we obtain
\be\label{ena}
a = - {1\over 2} + {1\over 2} \sqrt{1-{m^2R^4\over \ell_1^2}}
= - {1\over 2} + i\kappa\;, \quad\quad
\kappa = {1\over 2} \sqrt{{m^2R^4\over \ell_1^2} -1} \approx
{mR^2\over 2\ell_1} = {m\over 4\pi T_H}
\ee
where $T_H = {\ell_1\over 2\pi R^2}$ is the Hawking temperature.
As $u\to 1$, we obtain
\be
(1-u)^2 {d^2\Psi\over du^2} + (1-u) {d\Psi\over du} +
{m^2R^4\over 4\ell_1^2} \; (1-u)\Psi - {j(j+4)\over 4} \Psi = 0
\ee
Assuming $\Psi \sim (1-u)^b$, we obtain
$b = {j+4\over 2}$.
Now set
\be\label{tes1}
\Psi = A u^{-1/2 + i\kappa} \; (1-u)^{j/2+2} f(u)
\ee
Eq.~(\ref{tria1}) becomes
\be
(1-u)u {d^2f\over du^2} + [1+2i\kappa-(j+4+2i\kappa)u] {df\over du} -
{(j+3+2i\kappa)^2\over 4} \; f = 0
\ee
whose solution is the hypergeometric function
\be\label{pente1}
f(u) = F\left( (j+3)/2+i\kappa\; ,\; (j+3)/2+i\kappa\; ;\; 1+2i\kappa\; ;\; u
\right)
\ee
To obtain the behavior of $\Psi$ for large $r$, note that
\be\label{exi1}
F\left( (j+3)/2+i\kappa\; ,\; (j+3)/2+i\kappa\; ;\; 1+2i\kappa\; ;\; u
\right) = {\Gamma (1+2i\kappa)
\Gamma(j+2)\over (\Gamma((j+3)/2+i\kappa))^2}
 {1\over (1-u)^{j+2}} +\dots
\ee
where the dots represent terms that are regular in $1-u$.
Therefore, using Eqs.~(\ref{tes1}), (\ref{pente1}) and (\ref{exi1}), we arrive at
\be
\Psi \approx A \; {\Gamma (1+2i\kappa)
\Gamma(j+2)\over (\Gamma((j+3)/2+i\kappa))^2} \; \left( {r\over \ell_1}\right)^j
\ee 
Comparing with the asymptotic form~(\ref{tria}), we obtain
\be\label{exi}
A = {(\Gamma((j+3)/2+i\kappa))^2\over \Gamma (1+2i\kappa)}\;
{m^{j+2}\ell_1^j\over 2^{j+2}(j+1)!(j+2)!}
\ee
In the small $r$ limit, we have $u\approx r^2/\ell_1^2$ and Eq.~(\ref{tes1}) reads
\be
\Psi \approx A \left( {r\over \ell_1} \right)^{-1+2i\kappa}
\ee
The absorption coefficient, which is the ratio of the incoming flux
at $r\to 0$ to the incoming flux at $r\to \infty$, is
\be\label{epta1}
{\cal P} = {\Im (\lambda r^5 \Psi^\ast \Psi')|_{r\to 0}\over \Im (r^5 \Psi^\ast \Psi')|_{r\to\infty}}
= 4\pi \kappa \ell_1^4 |A|^2
= 4\pi \kappa \; {|\Gamma((j+3)/2+i\kappa)|^4\over |\Gamma (1+2i\kappa)|^2}\;
{m^{2j+4}\ell_1^{2j+4}\over 4^{j+2}((j+1)!(j+2)!)^2}
\ee
This is of the same form as the grey-body factors obtained in black-hole scattering~\cite{bib9b} for large $j$.
Indeed, comparing
\be
{\cal P} \sim |\Gamma((j+3)/2+i\kappa)|^4 = \left|\Gamma\left(
{j+3\over 2}+i\; {m\over 4\pi T_H}\right)\right|^4
\ee
with the general form of a grey-body factor~\cite{bib9b},
\be
{\cal P}_{b.h.} \sim \left|\Gamma\left(
{j+2\over 2}+i\; {m\over 4\pi T_L}\right)\right|^2\left|\Gamma\left(
{j+2\over 2}+i\; {m\over 4\pi T_R}\right)\right|^2
\ee
we see that we get contributions from both left- and right-moving modes at
temperatures $T_L=T_R=T_H$.

Eq.~(\ref{epta1}) also reproduces results derived earlier in the AdS (zero
temperature) limit. Indeed,
in the small temperature limit, we have $\kappa \to \infty$ and
\be
A \approx 
\sqrt\pi\; {i^{j+2} R^{2j+4}m^{2j+4}
\over 4^{j+2+i\kappa/2}\ell_1^2\; (j+1)!(j+2)!} \; {\Gamma (1/2+i\kappa)\over
\Gamma (1+i\kappa)}
\ee
for even $j$, where we used the Gamma function identities
\be\label{okto}
\Gamma (2x) = {1\over \sqrt{2\pi}} \; 2^{2x-1/2} \Gamma(x)\Gamma(x+1/2)
\quad\quad
\Gamma(x+1) = x \Gamma(x)
\ee
Since also $|\Gamma ({1\over 2} + i\kappa)|^2 = \pi / \cosh (\pi\kappa)$ and
$|\Gamma (1 + i\kappa)|^2 = \pi\kappa / \sinh (\pi\kappa)$, we obtain
\be
|A|^2 \approx { \pi R^{4j+8}m^{4j+8}
\over 4^{2j+4}\kappa \ell_1^4\; ((j+1)!(j+2)!)^2}
\ee
and so the absorption coefficient~(\ref{epta1}) becomes
\be\label{okto1}
{\cal P} = 4\pi \kappa \ell_1^4 |A|^2
\approx { \pi^2 R^{4j+8}m^{4j+8} \over 4^{2j+3} \; ((j+1)!(j+2)!)^2}
\ee
in agreement with earlier results~\cite{bib8a}.

\subsection{Two-component angular momentum}

Next, we turn to the case where two angular momentum quantum numbers are non-vanishing.
Let $\ell_1 = 0$, $\ell_2 = \ell_3$. Then the harmonic function $H$ (Eq.~(\ref{ena1})) becomes
\be
H = 1+ {R^4\over \left(1+{\ell_2^2\over r^2} \right)\zeta r^4}
= 1+ {R^4\over (r^2+\ell_2^2 )(r^2+\ell_2^2\sin^2\theta)}
\ee
The branes lie in the 4-sphere bounded by the surface $r=0$, {\em i.e.,}~they lie
inside a sphere of radius $\ell_2$ in the $y_1=y_2=0$ hyperplane.
Define ({\em cf.}~Eq.~(\ref{cootr}))
\be
y_\perp^2 = y_1^2+y_2^2 = r^2 \sin^2\theta \;, \quad\quad
y_{||}^2 = y_3^2+\dots + y_6^2 = (r^2+\ell_2^2)\cos^2\theta
\ee
As $r\to 0$, the harmonic function becomes
\be
H \approx 1+ {R^4r^2\over\ell_2^4y_\perp^2}
\ee
and so the density of D-branes is proportional to $r^2\to 0$. It follows that
there can be no D-branes in the interior of the four-sphere, therefore,
the D-branes are distributed on the surface of the four-sphere of radius $\ell_2$
defined by $r=0$. This distribution is uniform by symmetry, and the density is
$\sigma = {1\over 2\pi^2\ell_2^3}$.

The wave equation for small $r$ (Eq.~(\ref{duo1})) is
\be
{1\over r^3\sqrt\lambda} \left(
\lambda r^5 \Psi'\right)' + {R^4m^2\over \sqrt\lambda r^2} \Psi
- j(j+4) \Psi
= 0
\ee
where $\lambda = ( 1+\ell_2^2/ r^2)^2$.
Changing variables to $u=1/\sqrt\lambda$, we obtain
\be\label{ennia1}
u(1-u) {d^2\Psi\over du^2} + {d\Psi\over du} + {R^4m^2\over 4\ell_2^2\; u} \Psi
-{j(j+4)\over 4(1-u)} \Psi =0 
\ee
As $u\to 1$, we obtain
\be
(1-u)^2 {d^2\Psi\over du^2} + (1-u) {d\Psi\over du} +
{R^4m^2\over 4\ell_2^2} \; (1-u)\;\Psi - {j(j+4)\over 4} \;\Psi =0
\ee
Assuming $\Psi \sim (1-u)^b$, we find $b={j+4\over 2}$. As $u\to 0$, we obtain
\be
u^2 {d^2\Psi\over du^2} + u\; {d\Psi\over du} 
+ {R^4m^2\over 4\ell_2^2} \Psi -{j(j+4)\over 4} \; u\;\Psi = 0
\ee
Assuming $\Psi \sim u^a$, we find
\be
a = i \kappa\;, \quad\quad \kappa = {R^2m\over 2\ell_2}
\ee
Setting
\be
\Psi = A (1-u)^{j/2+2} \; u^{i\kappa} \; f(u)
\ee
the wave equation~(\ref{ennia1}) becomes
\be
u(1-u) f'' + ( 1+2i\kappa - (j+4+2i\kappa) u
) f' - ( j/2+1+i\kappa)(j/2+2+i\kappa) f = 0
\ee
whose solution is the hypergeometric function
\be
f(u) = F ( j/2+1+i\kappa\; , \; j/2+2+i\kappa \; ; \; 1+2i\kappa \; ; \; u)
\ee
In the large $r$ limit, we have $1-u \approx \ell_2^2 / r^2$, 
\be
f(\ell_2^2 / r^2) \approx \left( {\ell_2\over r} \right)^{-2j-4}\;
{\Gamma(1+2i\kappa) \Gamma(j+2)\over \Gamma(j/2+1+i\kappa)
\Gamma(j/2+2+i\kappa)}
\ee
and so
\be
\Psi \approx A \;{\Gamma(1+2i\kappa) \Gamma(j+2)\over \Gamma(j/2+1+i\kappa)
\Gamma(j/2+2+i\kappa)}\;
{r^j\over \ell_2^j}
\ee
Comparing with the asymptotic form~(\ref{tria}), we obtain
\be
A = {\Gamma(j/2+1+i\kappa)
\Gamma(j/2+2+i\kappa)\over \Gamma (1+2i\kappa)}\;
{m^{j+2}\ell_2^j\over 2^{j+2}(j+1)!(j+2)!}
\ee
In the small $r$ limit, we have $u\approx r^2/\ell_2^2$ and
\be
\Psi \approx A \left( {r\over \ell_1} \right)^{2i\kappa}
\ee
The absorption coefficient is ({\em cf.}~Eq.~(\ref{epta1}))
\be\label{epta2}
{\cal P} = 4\pi \kappa \ell_2^4 |A|^2
= 4\pi \kappa \; {|\Gamma(j/2+1+i\kappa)
\Gamma(j/2+2+i\kappa)|^2\over |\Gamma (1+2i\kappa)|^2}\;
{m^{2j+4}\ell_2^{2j+4}\over 4^{j+2}((j+1)!(j+2)!)^2}
\ee
in agreement with the one-component case (Eq.~(\ref{epta1})) and
with the same small-temperature ($\kappa\to\infty$) limit as before
(Eq.~(\ref{okto1})).

\subsection{Three-component angular momentum}

Next, we consider the case where all three angular momentum
quantum numbers are non-vanishing. This case is problematic, because of the
emergence of an infinite number of resonances. These problems arise because
the $r\to 0$ surface encloses a region of finite volume in the transverse space
spanned by the $y_i$ ($i=1,\dots,6$) coordinates~\cite{bib7aa}. In the cases previously considered,
when at most two angular momentum
quantum numbers were non-vanishing, the  $r\to 0$ surface enclosed a region
of zero measure in the transverse space. When this region has finite volume, the
possibility arises of the wave bouncing off of the branes (which are distributed on
the $r=0$ surface) an infinite number of times, hence the resonant behavior~\cite{bib10}.
We shall discuss two cases: {\em (a)} the spherically symmetric case ($\ell_1 =
\ell_2=\ell_3$), and {\em (b)} the `long needle' case ($\ell_2=\ell_3 \ll \ell_1$).
We will compute the absorption coefficients in both cases and show that they
are in agreement with our previous results. It should be emphasized that this
agreement is obtained when one performs the calculations in the Schwarzschild-like
coordinates~(\ref{miden}). The $y_i$ ($i=1,\dots,6$) coordinates span a larger
space which includes a region of finite volume surrounded by D-branes.
It appears that this region, which is disconnected from the region spanned by
the Schwarzschild-like coordinates~(\ref{miden}), should be excluded from physical
considerations.

First, let us consider the spherically symmetric case, $\ell_1 = \ell_2 =\ell_3$.
The wave equation~(\ref{duo1}) becomes
\be
{1\over r^3 \lambda^{1/3}}
\left(
\lambda r^5 \Psi'\right)' +
{R^4m^2\over r^2\lambda^{1/3}} \Psi
- j(j+4) \Psi
= 0
\ee
where
\be
\lambda = \left(1+{\ell_1^2\over r^2} \right)^3
\ee
Switching variables to $y = r\lambda^{1/6} = \sqrt{\ell_1^2+r^2}$ (note that
$y^2 = y_1^2+\dots +y_6^2$, due to Eq.~(\ref{cootr})), we obtain
\be\label{eq63}
{1\over y^3} \; (y^5\Psi')' + \left( {R^4m^2\over y^2} - j(j+4)\right) \Psi = 0
\ee
whose solution is
\be\label{tessera}
\Psi = A\; {1\over y^2} \; H_{j+2}^{(1)} \left( {R^2m\over y} \right)
\ee
We discarded the other solution, $H_{j+2}^{(2)}$, because we require
an incoming wave as $r\to 0$ (indeed, $H_{j+2}^{(1)} \sim y^{-3/2}
e^{iR^2m/y}$ for small $y$). In the large $r$ limit, we have $y \approx r$
and so
\be
\Psi \approx -i\; A\; {2^{j+2}\; (j+1)! \over R^{2j+4}
m^{j+2}} \; r^j
\ee
Comparing with the asymptotic form~(\ref{tria}), we obtain
\be\label{pente}
A = i\; {m^{2j+4} R^{2j+4} \over 4^{j+2} (j+1)! (j+2)!}
\ee
The absorption coefficient is
\be
{\cal P} = {16\pi R^2m\over\ell_1}\; |A|^2
\Im \left( H_{j+2}^{(1)\; \ast}\; H_{j+2}^{(1)\; '}\right)
= {16\pi R^2m\over\ell_1}\; {m^{4j+8} R^{4j+8} \over 4^{2j+4} ((j+1)! (j+2)!)^2}\; \Im \left( H_{j+2}^{(1)\; \ast}\; H_{j+2}^{(1)\; '}\right)
\ee
where the Bessel functions are evaluated at $2\kappa = R^2m /\ell_1$ (i.e., at $y=\ell_1$).
To compare with previous results, use
\be
H_{j+2}^{(1)} (2\kappa) = {(-i)^{j+5/2}\over\sqrt{\pi \kappa}} e^{2i\kappa}
\left( 1+ {i\over 4\kappa}\;
(j+5/2)(j+3/2) + \dots \right)
\ee
Using the Gamma function identities~(\ref{okto}), after some algebra we obtain
\be\label{eq68}
H_{j+2}^{(1)\;\ast} (2\kappa) = {2^{i\kappa} \; i\over \kappa^{j+2}} e^{-2i\kappa}
\; {\Gamma(j/2+5/4+i\kappa)\Gamma(j/2+7/4+i\kappa)\over
\Gamma(1+2i\kappa)} + \dots
\ee
and also
\be
H_{j+2}^{(1)\; '} (2\kappa) = {2^{-1-i\kappa} \; i\over \kappa^{j+3}}
e^{2i\kappa}
\; {\Gamma(j/2+7/4-i\kappa)\Gamma(j/2+9/4-i\kappa)\over
\Gamma(1-2i\kappa)} + \dots
\ee
Therefore, the absorption coefficient can be written as
\be\label{epta3}
{\cal P} = {4\pi \kappa \over ((j+1)! (j+2)!)^2}\; 
\; {|\Gamma(j/2+5/4+i\kappa)\Gamma(j/2+7/4+i\kappa)|^2\over
|\Gamma(1+2i\kappa)|^2} \;
\left( {m \ell_1\over 2}\right)^{2j+4}
\ee
in agreement with previous results (Eqs.~(\ref{epta1}) and (\ref{epta2})).

Complications arise when one continues into the $y<\ell_1$ region.
To do that, we need to assume a certain, spherically symmetric,
distribution of D-branes. Assuming
the space $y<\ell_1$ is empty, we obtain the wave equation
\be
{1\over y^3} \; (y^5\Psi')' + \left( {R^4m^2\over \ell_1^4}\; y^2
- j(j+4)\right) \Psi = 0
\ee
whose solution is
\be
\Psi = A'\; {1\over y^2} \; J_{j+2} \left( {R^2m\over \ell_1^2}\; y \right)
\ee
where we discarded the solution which is singular as $y\to 0$.

Also, we may now have an outgoing wave in the $y>\ell_1$ region, as well as
an incoming wave, so
Eq.~(\ref{tessera}) should be replaced by
\be
\Psi = A\; {1\over y^2} \; H_{j+2}^{(1)} \left( {R^2m\over y} \right)
+ B\; {1\over y^2} \; H_{j+2}^{(2)} \left( {R^2m\over y} \right)
\ee
and Eq.~(\ref{pente}) becomes
\be
A - B = i\; {m^{2j+4} R^{2j+4} \over 4^{j+2} (j+1)! (j+2)!}
\ee
Demanding continuity at $y=\ell_1$, we can obtain all coefficients, $A, B, A'$.
The system exhibits a resonant behavior when either the wavefunction or its derivative
vanishes on the D-brane (at $y=\ell_1$). In this case, we have either
$J_{j+2} (R^2m / \ell_1) = 0$ or $J_{j+2}' (R^2m / \ell_1) = 0$,
both of which have an infinite number of solutions, $m_n = x_n
\ell_1/R^2$, or $m_n = x_n'\ell_1/R^2$, where $J_{j+2} (x_n)= J_{j+2}' (x_n') =0$.

In hopes of shedding some light on this singular behavior, one may study the case
where two of the components of the angular momentum are small, {\em i.e.,}~let
us assume
\be
\ell_2 = \ell_3 \ll \ell_1 \lesssim R^2m
\ee
This is a small departure from the one-component case considered above and one
may hope to recover that solution in the limit $\ell_2\to 0$. Contrary to expectations,
we find that the resonances persist in the $\ell_2\to 0$ limit. Thus, even switching
on small components leads to a significant departure from the one-component case.
It appears that the region spanned by the Schwarzschild-like coordinates is more
physically relevant than the entire transverse space spanned by the coordinates
$y_i$ ($i=1,\dots,6$).

We will solve the wave equation by considering
two regimes, $r\gg \ell_2$ and $r\ll \ell_1 \lesssim R^2m$. When $r\gg \ell_2$, we can write
\be
{1\over r^3}
\left(\left( 1 + {\ell_1^2\over r^2} \right) r^5 \Psi'\right)' 
+ {R^4m^2\over r^2} \Psi- j(j+4) \Psi
= 0
\ee
which is identical to the one-component angular momentum case.
Therefore, the solution is
\be
\Psi = A\; u^{-1/2+i\kappa} (1-u)^{j/2+2} F (j/2+1+i\kappa\; ,\; j/2+1+i\kappa
\; ;\; 1+2i\kappa\; ; \; u)
\ee
where $u = 1/(1+\ell_1^2/r^2)$, A is given by Eq.~(\ref{exi}),
and $\kappa$ is given by Eq.~(\ref{ena}).
At small $r$, we have $u\sim r^2/\ell_1^2$, so
\be
\Psi \sim A\; \left( {r\over\ell_1} \right)^{-1+2i\kappa}
\ee
In the regime $r\ll \ell_1 \lesssim R^2m$, we can write the wave equation as
\be\label{duo}
{1\over r^3\left(1+{\ell_2^2\over r^2} \right)}
\left(\left(1+{\ell_2^2\over r^2} \right)^2 {\ell_1^2\over r^2}
r^5 \Psi'\right)' 
+{R^4m^2 \over r^2\left(1+{\ell_2^2\over r^2} \right)} \Psi- j(j+4) \Psi
= 0
\ee
Switching variables to $y = \sqrt{1+r^2/\ell_2^2}$, we obtain
\be\label{epta}
{1\over y} \; (y^3 \Psi')' + {R^4m^2\over \ell_1^2} \Psi - {j(j+4)\ell_2^2
\over \ell_1^2}
\; y^2 \; \Psi = 0
\ee
whose solution is
\be
\Psi \sim {1\over y} \; I_{2i\kappa} \left( {\sqrt{j(j+4)}\; \ell_2\over
\ell_1} \; y\right)
\ee
For sufficiently small $j$, the argument of the Bessel function is small,
so we can approximate
\be
\Psi \approx C\; y^{-1+2i\kappa}
\ee
At large $r$, we have $y\sim r/\ell_2$, so
\be
\Psi \sim C\; \left( {r\over \ell_2} \right)^{-1+2i\kappa}
\ee
Matching the two asymptotic forms, we obtain $C = A\; (\ell_2/\ell_1)^{-1+2i\kappa}$, and
\be
\Psi = A\; \left( {\ell_2^2+r^2\over \ell_1^2} \right)^{-1/2+i\kappa}
\ee
The absorption coefficient is found to be
\be
{\cal P} = 4\pi \kappa \ell_1^4 |A|^2 = 4\pi \kappa \; {|\Gamma((j+3)/2+i\kappa)|^4\over |\Gamma (1+2i\kappa)|^2}\;
{m^{2j+4}\ell_1^{2j+4}\over 4^{j+2}((j+1)!(j+2)!)^2}
\ee
in agreement with the one-component case (Eq.~(\ref{epta1})).

Again, complications arise when one considers the region $y<1$.
To illustrate the effect,
we shall only consider the case of an $s$-wave, $j=0$. In this case,
Eq.~(\ref{epta}) is the wave equation in cylindrical coordinates.
For higher $j$, we need to express the wavefunctions in terms of
cylindrical harmonics instead of spherical harmonics, and calculations become
increasingly involved for large $j$.
For $j=0$, we obtain for $y>1$,
\be
\Psi = C_1\; y^{-1+2i\kappa} +  C_2\; y^{-1-2i\kappa}
\ee
Assuming that the region $y<1$ is empty, the wave equation becomes
\be
{1\over y} \; (y^3\Psi')' + {R^4m^2\over \ell_1^2} \; y^2\; \Psi =0
\ee
whose solution is
\be
\Psi = D\; {1\over y} \sin \left( {R^2m\over \ell_1}\; y\right)
\ee
This exhibits resonant behavior when the wavefunction vanishes on the D-branes, at $y=1$. The resonances are at the points
\be
m_n = {n\pi \ell_1\over 2R^2}
\ee
Notice that in the limit $\ell_2 \to 0$, these resonances persist and it is not
clear how one may
recover the one-component solution discussed above.
However, if one excludes the cylindrical region $y<1$ enclosed by the D-branes from
physical considerations, then one obtains the same type of absorption coefficients
as in the rest of the cases (with one or two non-vanishing angular momentum
quantum numbers) and no resonances arise.

\section{A non-extremal D3-brane}\label{sec3}

When we try to go away from extremality, the wave equation becomes very complicated,
because the metric develops off-diagonal elements. Still, we would like to study
the more general background of non-extremal branes and study their limit as the
horizon shrinks to zero. Here we discuss the simplest case of all angular momentum
quantum numbers being zero. Setting $\ell_1=\ell_2=\ell_3=0$ in Eq.~(\ref{miden}),
the metric becomes
$$ds^2 = {1\over\sqrt H} \left( - (1-r_0^4/r^4) dt^2 + dx_1^2+dx_2^2+dx_3^2 \right)
+ \sqrt H {dr^2\over 1 - r_0^4/r^4}$$
\be
+ \sqrt H r^2 \left[ d\theta^2+ \sin^2\theta d\phi_1^2
+\cos^2\theta (d\psi^2 + \sin^2\psi d\phi_2^2
+ \cos^2\psi d\phi_3^2) \right]
\ee
where
\be
H = 1+ {R^4\over r^4} 
\ee
The radial part of the wave equation,
\be
\partial_A \sqrt{-g} g^{AB} \partial_B \Phi = 0
\ee
for fields that are independent of the angular variables $\psi$, $\phi_i$
($i=1,2,3$) and $\vec x$,
\be
\Phi = e^{i\omega t} \Psi (r) Y_j (\theta)
\ee
can be written as
\be
{1\over r^3}
\left(\left(1- {r_0^4\over r^4} \right) r^5 \Psi'\right)' 
+{r^2\omega^2\over (1-r_0^4/ r^4)} \; H \Psi - j(j+4) \Psi
= 0
\ee
We will solve this equation for wavelengths (horizons) of size much larger (smaller) than the AdS scale,
\be
r_0 \ll R \ll 1/\omega
\ee
We will also assume $r_0 \lesssim R^2\omega$ (so that we can
take the limit of the frequency being large compared to the temperature).

Away from the horizon, $r \gg R^2\omega$, we can replace $H$ by $1$
(by comparing its contribution to the $j(j+4)$ term). Therefore,
\be
{1\over r^3}
\left( r^5 \Psi'\right)' 
+r^2\omega^2 \; \Psi - j(j+4) \Psi
= 0
\ee
whose solution is
\be
\Psi = {1\over r^2} \; J_{j+2} (\omega r)
\ee
Next, we consider the region near the horizon, $r \ll R$.
In this case, $H\approx R^4/r^4$, and so
\be\label{eq99}
{1\over r^3}
\left(\left(1- {r_0^4\over r^4} \right) r^5 \Psi'\right)' 
+{R^4\omega^2\over r^2(1-r_0^4/ r^4)} \; \Psi - j(j+4) \Psi
= 0
\ee
To solve this equation, first we need to isolate the singularity at the
horizon. The wavefunction at the horizon behaves as $\Psi \sim (1- r_0^4/
r^4 )^{i\kappa}$. It is therefore convenient to define
\be
\Psi = A\; \left(1- {r_0^4\over r^4} \right)^{i\kappa} f(r)\quad\quad
\kappa = {R^2\omega\over 4r_0} = {\omega\over 4\pi T_H}
\ee
where $T_H = {r_0\over \pi R^2}$ is the Hawking temperature.
Then Eq.~(\ref{eq99}) becomes
\be\label{ennia}
r^2 \; \left(1- {r_0^4\over r^4} \right) \; f'' + r\; \left[
5 - (1-2i\kappa)\; {r_0^4\over r^4} \right] f' - j(j+4) f
= - {4R^4\omega^2\over r^2} \; {1+ {r_0^2\over r^2} + {r_0^4\over r^4} \over
1+ {r_0^2\over r^2} } \; f
\ee
The function $f$ has a regular limit as $r_0\to 0$ (as expected, since
we have already isolated the singularity in the wavefunction). Neglecting
higher-order corrections, we set $r_0 = 0$ in Eq.~(\ref{ennia}). The
result is ({\em cf.}~Eq.(\ref{eq63}))
\be
r^2 f'' + 5r f'
+ {R^4\omega^2\over r^2}\; f- j(j+4) f =0
\ee
whose solution is ({\em cf.}~Eq.(\ref{tessera}))
\be
f(r) = {1\over r^2} \; H_{j+2}^{(1)} \left( {R^2\omega\over r} \right)
\ee
In the large $r$ limit, we have
\be
\Psi \approx -i\; A\; {2^{j+2}\; (j+1)! \over R^{2j+4}
\omega^{j+2}} \; r^j
\ee
Comparing with the asymptotic form~(\ref{tria}), we obtain
\be
A = i\; {\omega^{2j+4} R^{2j+4} \over 4^{j+2} (j+1)! (j+2)!}
\ee
The absorption coefficient is
\be\label{eq106}
{\cal P} = 8\pi\kappa r_0^4 \; |A|^2 \; |f(r_0)|^2
\ee
Using the approximation~(\ref{eq68}), after some algebra we find that
for frequencies large compared to the temperature, the absorption coefficient~(\ref{eq106}) becomes
\be
{\cal P} \approx
{8\pi\kappa \over ((j+1)! (j+2)!)^2}
\; {|\Gamma(j/2+5/4+2i\kappa)\Gamma(j/2+7/4+2i\kappa)|^2\over
|\Gamma(1+4i\kappa)|^2}\; \left( {\omega r_0\over 2} \right)^{2j+4}
\ee
in line with the results we obtained in the extremal cases ({\em e.g.,}
Eq.~(\ref{epta3})), but at {\em half}~the Hawking temperature.

If we are allowed to speculate, we would like to note that this is reminiscent of
the case where there are two modes at temperatures $T_1$ and $T_2$, and the Hawking
temperature is given by
$2/T_H = 1/T_1+1/T_2$~\cite{bib9b}. If $T_2\to \infty$, then $T_1 = T_H/2$. Thus, when we
go off extremality, it seems that the number of degrees of freedom doubles with the
extra degrees living in a very hot bath. Of course, all this needs to be taken with
a grain of salt, since away from extremality supersymmetry is broken and
there is no guarantee that the
supergravity analysis is in any way dual to the superconformal field theory
on D-branes. Still, it is intriguing that similar results are obtained for both
non-extremal and extremal supergravity backgrounds.

\section{Conclusions}\label{sec4}

We discussed the absorption of scalar fields by a distribution of D3-branes
in the extremal limit. These distributions are obtained as limiting cases
of spinning branes which are solutions of the full non-linear supergravity
field equations~\cite{bib7a,bib7aa,bib7b,bib7c}.
In the extremal limit, the branes are no longer spinning, but they settle
into a state which is distinct from the AdS limit and is characterized by
angular momentum quantum numbers $\ell_i$ ($i = 1,2,3$). The AdS limit
is obtained when these quantum
numbers approach zero. This set of supergravity solutions is dual to the
Coulomb branch of the ${\cal N} = 4$ four-dimensional $SU(N)$ super Yang Mills
theory, which is a superconformal field theory.
We solved the wave equation for scalar fields in the respective supergravity
backgrounds and computed the absorption coefficient in each case. We found
that the absorption coefficients exhibited a universal behavior as functions
of the angular momentum quantum number of the partial wave and  the
Hawking temperature. This functional dependence is of the same form as the
grey-body factors associated with black-hole scattering~\cite{bib9b}.

We also discussed the problematic case of a spherically symmetric distribution
of D-branes~\cite{bib10}. This is an example of the more general case where
the D-branes are distributed on a surface that divides the transverse
space, defined by coordinates $y_i$ ($i=1,\dots,6$) (Eq.~(\ref{deka})) into
two distinct regions. This occurs when all three quantum numbers $\ell_i$ ($i = 1,2,3$) are non-vanishing~\cite{bib7aa}.
The wave can then bounce off of the D-branes an infinite number of times
and when this happens, one obtains a resonance. We obtained these
resonances for both a spherically symmetric and
a `long-needle' distribution of branes.
They present a puzzle as one approaches the AdS limit. However, if one uses
the Schwarzschild-like coordinates~(\ref{miden}) instead, which only
cover the region outside the D-brane shell, thereby excluding the inner
region from physical considerations, we showed that one obtains the
same form for the absorption coefficients as in the rest of the cases.
It seems that the Schwarzschild-like coordinates~(\ref{miden}) are more
appropriate physically than the $y_i$ ($i=1,\dots,6$) coordinates.

Finally, we went off extremality and solved the wave equation in the
background of a brane with a finite event horizon. Here, too, we obtained
the same form for the absorption coefficients, albeit at {\em half} the
Hawking temperature. We speculated that off extremality, new degrees of freedom
enter which must live in a hot bath. Of course, since supersymmetry is broken,
there is no guarantee that a duality exists between supergravity and
superconformal field theory on D3-branes. Yet, it is intriguing that similar
results are obtained in both the extremal and the non-extremal cases, albeit
with a twist.

It would be interesting to extend this analysis to more general supergravity
backgrounds. This is important for a better understanding of the
maximally supersymmetric AdS limit and its thermodynamic properties.
It will shed more light on the interesting issue of the
AdS/CFT correspondence.

\end{document}